\documentclass[twocolumn,showpacs,amsmath,amssymb,superscriptaddress,floatfix]{revtex4-1}
\usepackage{graphicx}
\usepackage{dcolumn}
\usepackage{bm}
\begin{document}

\title{Commensurate vs incommensurate charge ordering near the superconducting dome in Ir$_{1-x}$Pt$_x$Te$_2$ revealed by resonant x-ray scattering}

\author{K. Takubo}
\email{ktakubo@issp.u-tokyo.ac.jp}
\affiliation{Institute for Solid State Physics, University of Tokyo, Kashiwa, Chiba 277-8581, Japan}
\author{K. Yamamoto}
\affiliation{Institute for Solid State Physics, University of Tokyo, Kashiwa, Chiba 277-8581, Japan}
\author{Y. Hirata}
\affiliation{Institute for Solid State Physics, University of Tokyo, Kashiwa, Chiba 277-8581, Japan}
\author{H. Wadati}
\affiliation{Institute for Solid State Physics, University of Tokyo, Kashiwa, Chiba 277-8581, Japan}
\author{T. Mizokawa}
\affiliation{Department of Applied Physics, Waseda University, Okubo, Tokyo 169-8555, Japan}
\author{R. Sutarto}
\affiliation{Canadian Light Source, University of Saskatchewan, Saskatoon, Saskatchewan S7N 2V3, Canada} 
\author{\mbox{F. He}}
\affiliation{Canadian Light Source, University of Saskatchewan, Saskatoon, Saskatchewan S7N 2V3, Canada}
\author{K. Ishii}
\affiliation{Synchrotron Radiation Research Center, National Institutes for Quantum and Radiological Science and Technology, Hyogo 679-5148, Japan}
\author{\mbox{Y. Yamasaki}}
\affiliation{Research and Services Division of Materials Data and Integrated System (MaDIS), National Institute for Materials Science (NIMS), Tsukuba 305-0047, Japan}
\author{H. Nakao}
\affiliation{Institute of Materials Structure Science, High Energy Accelerator Research Organization, Tsukuba, Ibaraki 305-0801, Japan}
\author{Y. Murakami}
\affiliation{Institute of Materials Structure Science, High Energy Accelerator Research Organization, Tsukuba, Ibaraki 305-0801, Japan}
\author{G. Matsuo}
\affiliation{Research Institute for Interdisciplinary Science, Okayama University, Okayama 700-8530, Japan}
\author{H. Ishii}
\affiliation{Research Institute for Interdisciplinary Science, Okayama University, Okayama 700-8530, Japan}
\author{M. Kobayashi}
\affiliation{Research Institute for Interdisciplinary Science, Okayama University, Okayama 700-8530, Japan}
\author{K. Kudo}
\affiliation{Research Institute for Interdisciplinary Science, Okayama University, Okayama 700-8530, Japan}
\author{M. Nohara}
\affiliation{Research Institute for Interdisciplinary Science, Okayama University, Okayama 700-8530, Japan}

\date{\today}

\begin{abstract}
The electronic-structural modulations of Ir$_{1-x}$Pt$_x$Te$_2$ (0 $\leqq x\leqq$ 0.12) have been examined by resonant elastic x-ray scattering (REXS) and resonant inelastic x-ray scattering (RIXS) techniques at both the Ir and Te edges.
Charge-density-wave-like superstructure with wave vectors of $\mathbf{Q}$=(1/5 0 $-$1/5), (1/8 0 $-$1/8), and (1/6 0 $-$1/6) are observed on the same sample of IrTe$_2$ at the lowest temperature, the patterns of which are controlled by the cooling speeds.
In contrast, superstructures around $\mathbf{Q}$=(1/5 0 $-$1/5) are observed for doped samples (0.02 $\leqq x\leqq$ 0.05).
The superstructure reflections persist to higher Pt substitution than previously assumed, demonstrating that a charge density wave (CDW) can coexists with superconductivity.
The analysis of the energy-dependent REXS and RIXS lineshape reveals the importance of the Te 5$p$ state rather than the Ir 5$d$ states in the formation of the spatial modulation of these systems.
The phase diagram re-examined in this work suggests that the CDW incommensurability may correlate the emergence of superconducting states as-like Cu$_x$TiSe$_2$ and Li$_x$TaS$_2$.

\end{abstract}

\maketitle

\section{Introduction}

The interplay between spin-orbit coupling and the Coulomb interaction led to a renaissance in the study of transition-metal compounds,
because it can lead to novel superconductivity competing with charge ordering of spin-orbit Mott states as in high-$T_\mathrm{C}$ superconductors \cite{Mizokawa01,Kim08,Comin13,Comin15,Achkar13,Tranquada95,Ghiringhelli12}.
When the electronic states are localized as in the Mott state, a charge-ordered-type modulation appears owing 
to intersite Coulomb interactions.
Moreover, a charge modulation can be induced by the Peierls instability --- the so-called charge density wave (CDW).
In a system with heavy elements such as 5$d$ transition-metals, the large spin-orbit interaction can stabilize the localized spin-orbit Mott state, as observed in the 5$d$ transition-metal compounds, and then the charge-ordered Mott state leads to a novel framework of the interplay \cite{Ko15,Kim08}.

A CDW-like structural transition was reported in the 5$d$ transition-metal chalcogenide IrTe$_2$ at $T_s\sim280$ K \cite{Matsumoto99,Jobic92}.
This has attracted great interest because of the recent discovery of superconductivity in Pt- and Pd-substituted or intercalated compounds \cite{Pyon12,Pyon13,Yang12,Qi12,Kamitani13,Kudo13}.
With increasing Pt-substitution, the structural transition is suppressed and a superconducting dome appears 
in the region of 0.04 $\leqq x\leqq$ 0.12, indicating similar diagrams those of other unconventional superconductors.
Although numerous studies have followed these initial works, consensus about the mechanism for this structural transition is still lacking.
The phase transition of IrTe$_2$ ($x=0.0$) is accompanied by the emergence of a superstructure lattice modulation \cite{Yang12}, with wave vector $\mathbf{Q}_{1/5}$=(1/5 0 $-$1/5) as expressed in reciprocal lattice units in trigonal notation, which is illustrated in Fig. 1.
The main elements are the Ir-Ir dimerization along the $a$-axis with period 5$a$, and the consequent distortion of the triangular Ir sublattice in the $a$-$b$ plane, occurring together with a trigonal-to-triclinic symmetry reduction.
The Ir-Ir dimerization stabilizes a unique stripe-like order, with stripes running along the $b$-axis, as indicated by x-ray diffraction \cite{Cao13,Pascut14B,Pascut13,Toriyama13}, extended x-ray absorption fine structure \cite{Joseph13} and resonant x-ray scattering \cite{Takubo14} studies.
Since in IrTe$_2$ the formal valence of Ir is +4, the Ir 5$d$ electrons with $t_{2g}$ configuration are the closest to the chemical potential, and they are thus expected to play a central role in the CDW.
However, photoemission and optical studies have shown that the charge-transfer energy in IrTe$_2$ is close to zero, and that the Te 5$p$ states are also important for the low-energy physics \cite{Ootsuki12,Fang13,Qian13}.
In addition, recent x-ray diffraction (XRD) \cite{Pascut14B,Ko15} and scanning tunneling microscopy (STM) \cite{Hsu13} experiments revealed a further step-wise charge-ordering transition from $\mathbf{Q}_{1/5}$ to $\mathbf{Q}_{1/8}=$(1/8 0 $-$1/8) and/or $\mathbf{Q}_{1/11}=$(1/11 0 $-$1/11) below $T\sim$ 200 K, while many studies reported that the $\mathbf{Q}_{1/5}$-type superstructures survived at their lowest temperatures \cite{Toriyama13,Cao13,Takubo14}.
Results from these studies also suggested that a CDW occurs at both of the Ir and Te sites at least near its surface region.
Furthermore, in some studies, differences of the electronic state between surface and bulk states for IrTe$_2$ have been reported.
In another recent STM study for non-substituted IrTe$_2$, a superconducting domain coexisting with very complex charge-ordering structures that only exist in the surface region were reported \cite{Kim16}.
An exotic one-dimensional surface state was also observed on angular-resolved photoemission spectra by Ootsuki \textit{et al} \cite{Ootsuki13}. 
\begin{figure}[t!]
	\includegraphics[width=1\linewidth]{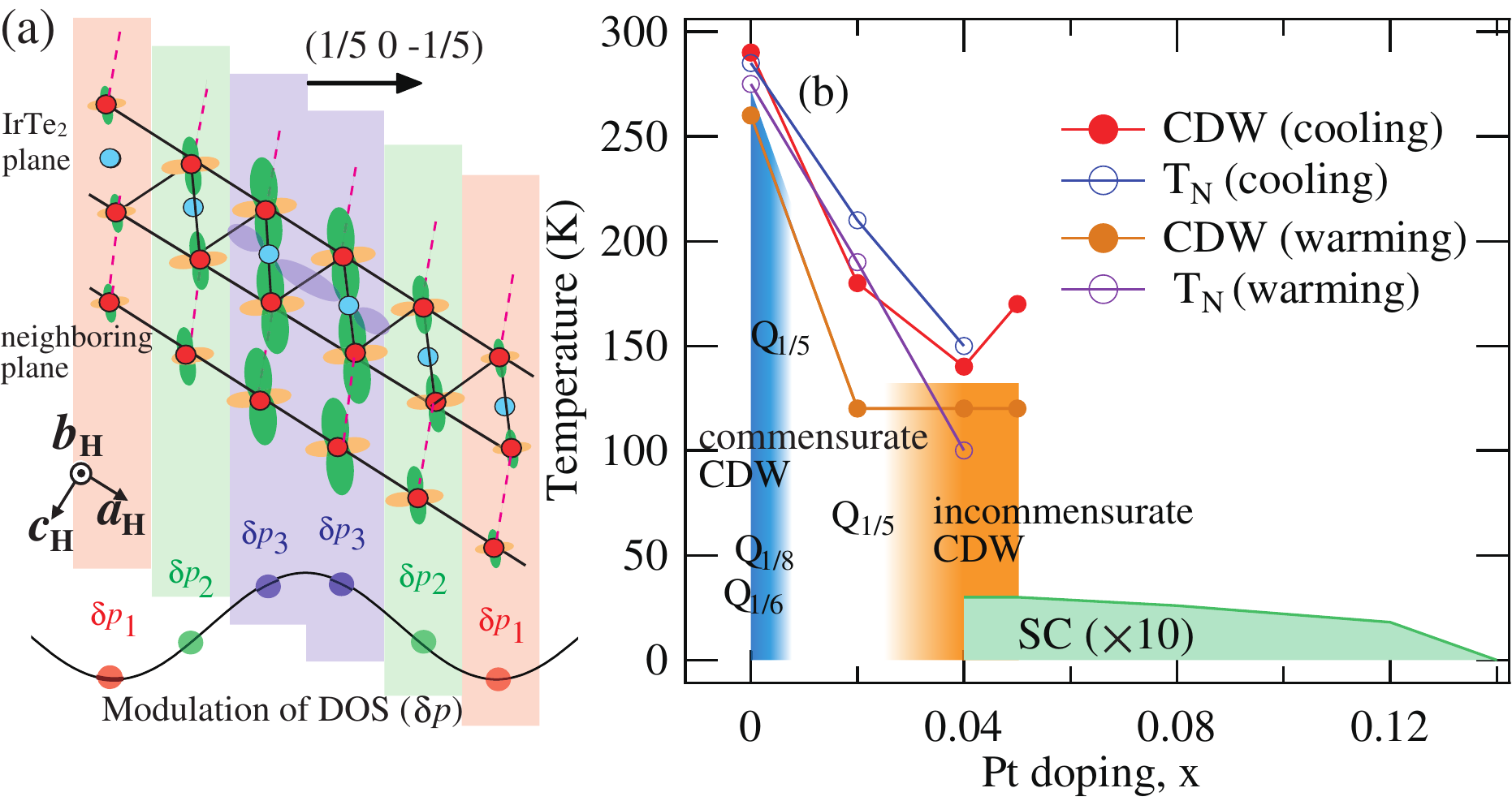}%
	\caption{(color online).
		(a) Superstructure modulation of IrTe$_2$ with wave vector $\mathbf{Q}_{1/5}$=(1/5 0 $-$1/5), as expressed in reciprocal lattice units in tetragonal notation \cite{Takubo14}.
		The modulation of the density of states (DOS), highlighting an Ir-Ir dimerization, is shown at the bottom.
		(b) Phase diagram of Ir$_{1-x}$Pt$_{x}$Te$_{2}$. Points denoted as CDW were obtained in this study. $T_N$ and the superconducting region are from Ref. \onlinecite{Pyon13}.
	}
\end{figure}

Moreover, a complex picture can be assumed for the relationship between such elusive CDW orders and superconducting orders in Pt-substituted Ir$_{1-x}$Pt$_x$Te$_2$.
Because the phase digram of Ir$_{1-x}$Pt$_x$Te$_2$ exhibits a diagram similar to those of other unconventional superconductors,
the idea of a quantum critical point inside the superconducting dome can be considered \cite{Pyon12}.
However, it has been observed that many other transition-metal chalcogenides exhibit the coexistence of superconductivity with incommensurate CDW orders, where disorder effects and incommensuration of the CDWs suggested to be more important \cite{Morosan06,Li15,Joe14,Liu13,Sipos08,Kogar17}. 
In the intercalated 1$T$-TaS$_2$ and 1$T$-TiSe$_2$ systems, the superconductivity only coexists with incommensurate CDWs, although their mother compounds exhibit commensurate CDWs \cite{Sipos08,Liu13,Kogar17}.

To revisit the superstructures in Ir$_{1-x}$Pt$_x$Te$_2$ and to clarify the relation between superconductivity and superstructural modulation in the bulk region,
we studied the spatial ordering of electronic states by means of bulk-sensitive methods: resonant elastic and inelastic x-ray scattering at edges of both Ir and Te.
As a result, superstructure peaks with wave vectors of $\mathbf{Q}$=(1/5 0 $-$1/5), (1/8 0 $-$1/8), and (1/6 0 $-$1/6) are found on IrTe$_2$ ($x$=0.0) at the lowest temperature, which are governed by the cooling speeds.
In contrast, the incommensurate ordering peaks around $\mathbf{Q}$=(1/5 0 $-$1/5) are observed for doped samples of 0.02 $< x \leqq$0.05 at low temperature, suggesting that CDWs can coexist with superconductivity for $x$=0.05. 
The resonant elastic and inelastic x-ray scattering results at the Ir and Te edges emphasize the importance of the Te 5$p$ states rather than Ir 5$d$ states in the stripe-like ordering formation in these systems.

\section{Experiment}

Single-crystal samples of Ir$_{1-x}$Pt$_x$Te$_2$ (0 $\leqq x\leqq$ 0.12) were prepared using a self-flux method \cite{Fang13,Pyon13,Kudo13}.
The cleaved  (001) planes were used for all the scattering experiments. 
Resonant elastic x-ray scattering (REXS) at the Ir $L_{3}$ (2$p$ $\rightarrow$ 5$d$) absorption edge in the hard x-ray region were performed at Photon Factory's BL-4C.
REXS at the Te $L_1$ (2$s$ $\rightarrow$ 5$p$) absorption edge were conducted at BL-22XU of SPring-8.
The polarization of incident x ray was perpendicular to the scattering plane.
The samples were mounted so that [100] and [001] directions were in the scattering plane,
although it was confirmed that the REXS spectra barely show the azimuthal dependence.
Here, the reciprocal space indices ($h$ $k$ $l$) refer to the high-temperature trigonal unit cell.
The x-ray absorption spectra (XAS) at the Ir $L_3$ and Te $L_1$ edges were recorded by their fluorescence.

On the other hand, resonant inelastic x-ray scattering (RIXS) at the Ir $L_3$ edges were carried out at BL-11XU of SPring-8 \cite{Ishii13}.
Incident x rays were monochromatized by a double-crystal Si(111) monochromator and a secondary Si(844) channel-cut monochromator.
Horizontally scattered x rays were analyzed in energy by a spherically diced and bent Si(844) crystal.
The total energy resolution was about 70 meV.
The spectra for horizontally polarized incident x-rays were recorded near $2\theta\sim$86 degrees so that elastic scattering was reduced [See Fig. 5(d)].
The sample was also mounted so that the [100] and [001] directions span the scattering plane.

REXS at the Te $M_{4,5}$ (3$d$ $\rightarrow$ 5$p$) edge in the soft x-ray region were performed at the REIXS beamline of Canadian Light Source \cite{Hawthorn_RevSci}.
Single crystals were cleaved in vacuum to minimize surface contamination effects.
The cleaved (001) plane was oriented at $\sim$54 degrees from the scattering plane in order to perform REXS
measurements in the $Q$=($h$ 0 $-h$) plane.
The polarization of the incident x ray was perpendicular to the scattering plane.
XAS at the Te-${M}_{4,5}$ edges were recorded in the total electron yield (TEY) modes.
The XAS results using the TEY mode showed no noticeable difference with respect to spectra acquired in total fluorescence yield mode.

\section{Results and Discussion}

\subsection{Superstructures in IrTe$_2$ ($x$=0.0)}

Figure 2 shows XRD along (0 0 $-$4) to (1 0 $-$5) through the superstructure peaks for IrTe$_2$ ($x$=0.0) taken with $hv=11.15$ keV which is below the energy of the Ir $L_3$ absorption.
The periods of the superstructures at low temperature strongly depend on the cooling protocols.
As for the results of $x$=0.0 shown in Fig. 2, the sample temperatures were continuously ramped down from $T$=300 K to 220 K with various tuned cooling rates and once XRD was measured at $T$=220 K.
Then the samples were cooling down again to $T$=10 K with the same speeds.
These measurements were conducted for a fresh sample each time grown in a single batch.
The measurements at $T$=10 and 220 K took for 1 $\sim$ 2 hours, including the time for alignment of the sample axes.
When the cooling rate was set to 2 K/min,
CDW-like superstructures with ${\mathbf Q}_{1/8}$=(1/8 0 $-$1/8) emerged at low temperature ($T$=10 K) as shown in Fig. 2 (b).
While only the Bragg peaks were observable at $T$=300 K [Fig.2(b)-(d)], 
the ${\mathbf Q}_{1/5}$=(1/5 0 $-$1/5)-type ordering peaks appeared at $T$=220K (below $T_s\sim$280 K).
Then, the modulation was subsequently changed to ${\mathbf Q}_{1/8}$-type below $T\sim$200 K,
although the ${\mathbf Q}_{1/5}$-type superstructures were reported for IrTe$_2$ in the previous studies for these samples at the lowest temperature \cite{Takubo14, Toriyama13, Pyon13}. 
Although some diffuse scatterings were observed along the ($H$ 0 $-H$) reciprocal axis, the ${\mathbf Q}_{1/8}$-type ordering stabilized at $T$=10 K.
This observation is similar to the step-wise charge-ordering cascade reported by Ko \textit{et al.} for samples synthesized by other groups \cite{Ko15}.
Upon heating, the ${\mathbf Q}_{1/8}$ peaks remain up to $T\sim290$ K and this phase directly transits to the high-temperature phase \cite{Ko15}. 
When the cooling speed was set to 4 K/min, however,
the step-wise cascade transition to ${\mathbf Q}_{1/8}$ was not observed and superstructures with ${\mathbf Q}_{1/5}$ were stabilized even at the lowest temperature of $T=10$ K as shown in Fig. 2(a),
which is consistent with the previous reports for the same samples \cite{Takubo14, Toriyama13}.
However, the coexistence of ${\mathbf Q}_{1/5}$ and ${\mathbf Q}_{1/8}$ could be obtained with a medium-cooling speed of 2.5 K/min.
Finally, superstructures around ${\mathbf Q}_{1/6}$=(1/6 0 $-$1/6) coexisting with weak ${\mathbf Q}_{1/8}$ peaks appeared with the fastest cooling speed above 5 K/min.
${\mathbf Q}_{1/6}$ ordering for IrTe$_2$ has not ever been reported previously, while similar superstructures were obtained for IrTe$_{2-x}$Se$_x$ ($x>0.3$) \cite{Oh13,Pascut14B}.
These observation strongly supports a scenario of anionic depolymerization transition, as suggested by Oh \textit{et al.} \cite{Oh13}, 
where depolymerization-polymerization occur between the anionic Te-Te bonds across the transition.
The covalence of the Te-Te bonds is partially lost across the transition, depolymerizing the Te-Te networks and leading to the diversity of superstructures.

The superstructures appear along one direction ($h$ 0 $-$4$-h$) of the triangular lattices
and are not observable along other two directions of (0 $k$ $-$4$-k$) and ($h$ $-h$ $-$4$-h$) as plotted in Fig. 2(c), indicating formation of a single-domain structure with an x-ray spot size of $\sim$ 1 mm$^2$. 
CDW distortion seemed to occur along one side of the triangular axis in these hard x-ray experiments,
although multi domain structures were reported in previous soft x-ray and low-energy electron diffraction experiments \cite{Takubo14}.
The evolution of the domain structures and superstructures seems to depend on the sample condition.
Figure 2(d) shows the patterns at $T$=220 K of the first and second cooling-warming cycles taken for the same sample,
which are normalized by the intensity of the Bragg peak of (004).
The superstructure intensities at the second cooling attempt were 10$^{-2}$ order of magnitude smaller than those at the first attempt.    
Since the structural transition of IrTe$_2$ across $T_s$ is very steep and can cause cracks in the crystal,
long-range ordering would be weakened after the cooling cycle.

It should be noted that the pattern of the superstructures could also depend on many other subtle conditions.
For example, the cracks mentioned above seem to change the thermal condition.
We also tried the repeated --- cooling and warming --- measurements on a single sample but did not observe a clear reproducibility.
However, ${\mathbf Q}_{1/8}$ superlattices were obtained in many cases, even when the cooling speeds were set to above 4 K/min.
The thermal conductance from cryostat would become worse by the cracks and this tendency would still be consistent with the scenario of the cooling-rate dependence.
In addition, the drift of the observations with time was also appeared if the temperature was fixed at the range between $T$=80 and 200 K.
The diffuse scattering along the ($H$ 0 $-H$) axis were sometimes observed at this temperature range (not shown). However, these drifts seemed to be very slow taking several hours, and a clear reproducibility has not been obtained. A further study will be needed in order to clarify this point.

\begin{figure*}[ht]
	\includegraphics[width=1\linewidth]{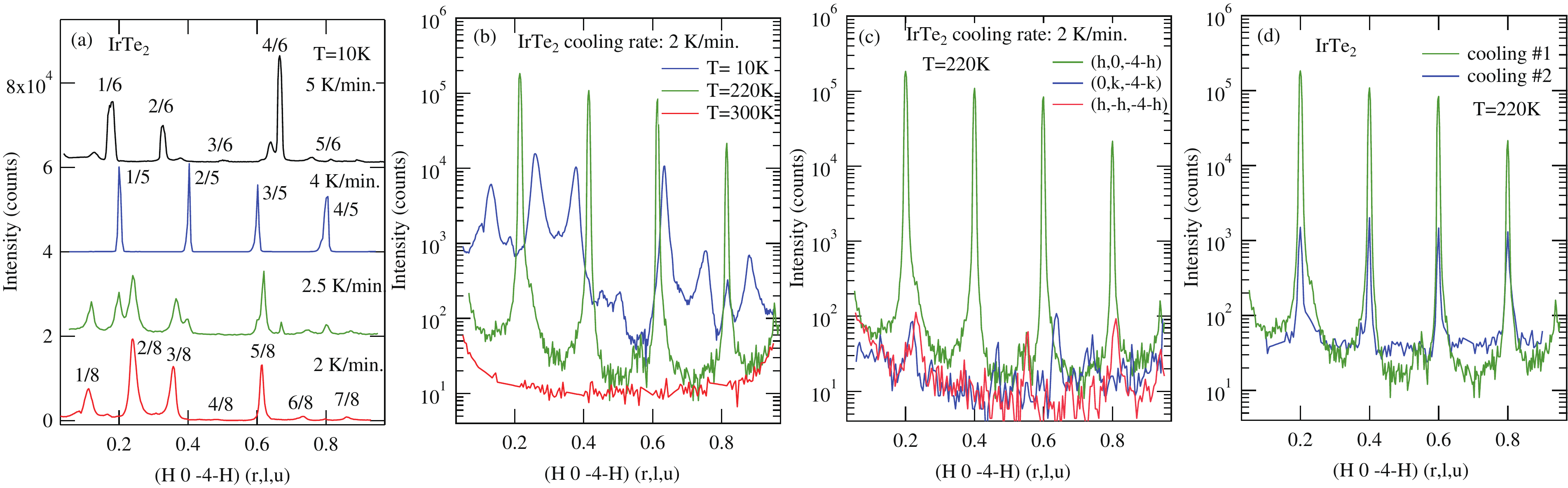}
	\caption{
		(color online.) ($H$ 0 $L$) scan through the superstructure peaks for IrTe$_2$ ($x$=0.0) taken with $hv=11.15$ keV.
		(a) Superstructure peaks at $T$=10 K obtained with various cooling rates: 2, 2.5, 4, and 5 K/min.
		(b) Temperature dependence of the superstructure peaks. The cooling rate was set to 2 K/min. 
		(c) ($H$ 0 $L$) scan along the 3 different crystal axes of ($h$ 0 $-$4$-h$), (0 $k$ $-$4$-k$), and  ($h$ $-h$ $-$4$-h$) at $T$=220 K.
		(d) Superstructure peaks at $T$=220 K in the first and second cooling-warming cycles for the same sample. The cooling rates were 2 K/min.
	}
\end{figure*}

\subsection{Doping dependence of superstructure}

Next, to clarify the relation between the CDW-like structural modulation and superconductivity,
the evolution of the superstructure peaks with Pt doping is examined by XRD for Ir$_{1-x}$Pt$_x$Te$_2$.
As for the results shown in Fig. 3, the averaged cooling rates were set to $\sim$ 2.5 K/min.
The measurements took $\sim$ 10 min. at each temperature.
The modulation periods only slightly depend on the cooling speeds, and cascade transition to ${\mathbf Q}_{1/8}$ were not observed for the doped samples (0.02 $\leqq x\leqq$ 0.05).
As can be seen, the CDW-like superstructures of {\bf Q}$_{1/5}$ are observed for $x$=0.02 below $T\sim$ 140 K.
In addition, the superstructures around ${\mathbf Q}_{1/5}$ are also found on $x$=0.04 and 0.05 samples at low temperature.
Although $x$=0.04 show some sample dependence of the structural and superconducting transition temperatures which may originate from its inhomogeneity \cite{Pyon13,Kudo13}, all the $x$=0.05 samples including the batch used in this study show the superconductivity and do not show any anomaly at $T_s\sim$140 K on the macroscopic conductivity and magnetization measurements [see Fig. 1(b)].
The superstructures for $x$=0.05 are 10$^{-2}$ order of magnitude smaller than that for $x$=0.02 but certainly observable around ${\mathbf Q}_{1/5}$.
4 pieces of $x$=0.05 crystals were investigated and similar superlattices were found on all the pieces.
The microscopic phase separation of the majority superconducting and minority charge ordered domains
may occur in the $x$=0.05 samples.
On the other hand, the superstructures disappears in $x$=0.08 and 0.12 [Fig. 3(b)].
CDW with ${\mathbf Q}_{1/5}$ seems to persist to higher doping level ($x\leqq0.05$) than previously thought and coexists with the superconducting state. 
Furthermore, CDW incomensuration is found along the ($H$ 0 $-H$) direction for $x$=0.04 and 0.05 near the superconducting dome.
Although the superstructures are perfectly commensurate with the lattice for $x$=0.0 and 0.02, the peaks for $x$=0.04 and 0.05 shift to the lower-$H$ side.
The peak widths also broaden as the doping level $x$ increases.
These observations are very similar to the results obtained for CDW in 1$T$-TaS$_2$ \cite{Sipos08} and 1$T$-TiSe$_2$ \cite{Kogar17,Joe14} systems,
where the incommensuration of CDWs also coincides with the onset of superconductivity.
Both electron-phonon and electron-hole couplings have been suggested to play significant roles in of these systems \cite{Kogar17}.
Therefore, similar mechanisms may also be important in driving the superconductivity of Ir$_{1-x}$Pt$_x$Te$_2$.
Another possibility for driving the incommensurability of $x$=0.04 and 0.05 is the coexistence of ${\mathbf Q}_{1/5}$- and ${\mathbf Q}_{1/8}$- type domains in the microscopic region, as observed in the STM studies for IrTe$_2$ where various kinds of ordered domains coexists on nanometer scales \cite{Hsu13, Kim16}.
Since the superconducting transition temperature at $x$=0.05 is similar to those for $x>$ 0.05 without CDW, it is naturally speculated that the superconducting phase at $x$=0.05 is different from the incommensurate CDW phase observed at $x$=0.04 and 0.05. However, the incommensurability of CDW shows that the phase separated $x$=0.05 state is not a mere mixture of the commensurate CDW phase at $x$=0.0 and the superconducting phase for $x>$0.05. In this sense, the incommensurability of CDW is associated with the emergence of superconductivity.

\begin{figure}[ht]
	\includegraphics[width=1\linewidth]{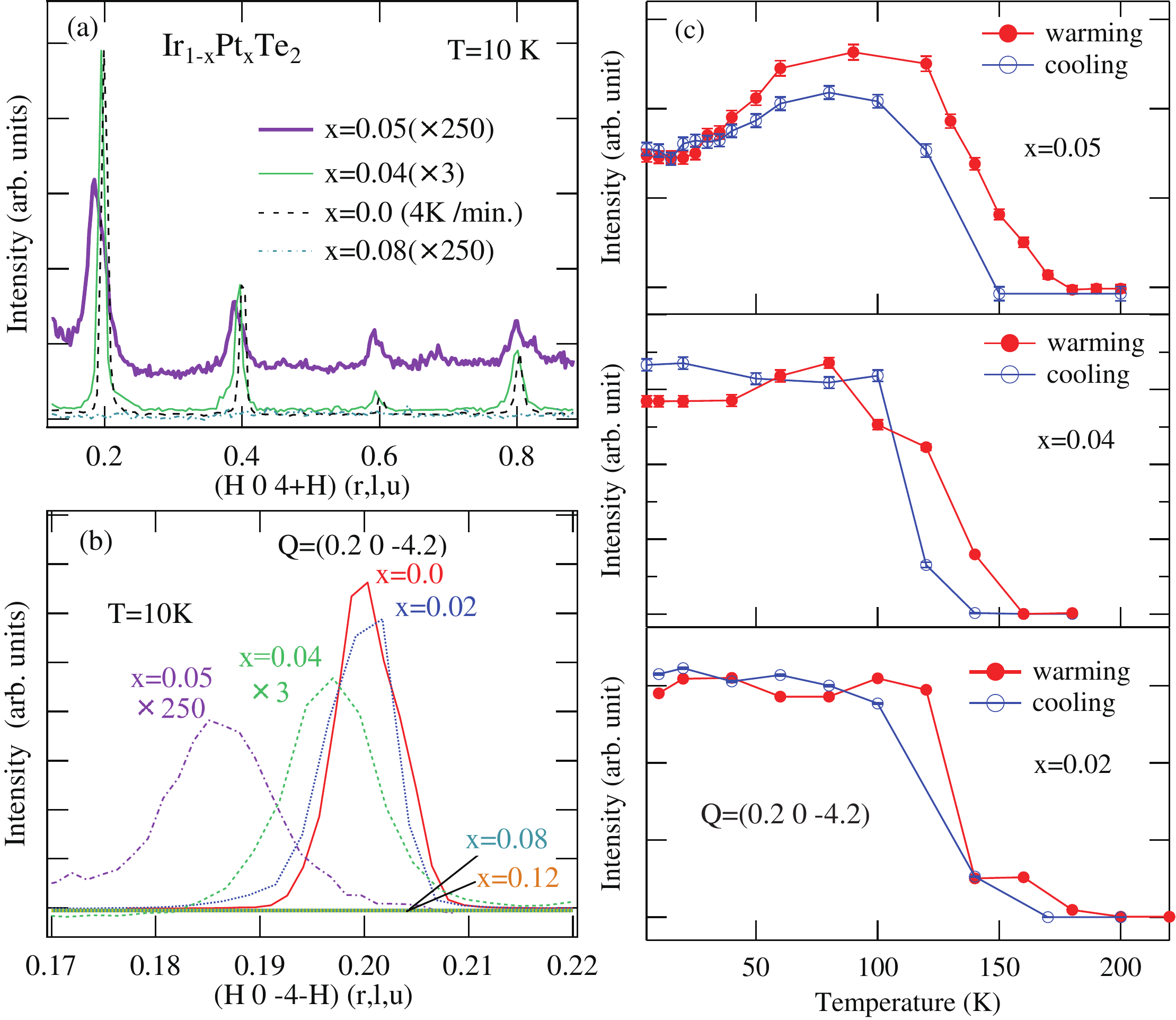}
	\caption{
		(color online.) (a) ($H$ 0 $L$) scan through the superstructure peaks for Ir$_{1-x}$Pt$_x$Te$_2$ ($x$=0.0, 0.04 and 0.05) taken with $hv=11.15$ keV.
		(b) ($H$ 0 $L$) scan through the superstructure peak of $\mathbf{Q}$=(0.2 0 -4.2) for Ir$_{1-x}$Pt$_x$Te$_2$ ($0.0 \leqq $x$ \leqq 0.12$). 
		(c) Temperature dependence around $\mathbf{Q}$=(0.2 0 -4.2) peaks for $x$=0.05 (top), $x$=0.04 (middle), and $x$=0.02 (bottom),
		which were evaluated as the sum of the counts over the whole peak of (0.2 0 $-$4.2). 
	}
\end{figure}

Figure 3(c) shows the detailed temperature dependence of the superstructure intensities around ${\mathbf Q}_{1/5}$ for $x$=0.02, 0.04 and 0.05, measured across $T_s$ during both cooling and warming cycles.
The cooling-warming rates were set to $\sim$4 K/min. 
The signals show sharp onsets having some hysteretic behaviors at $T_s\sim$ 150 K for $x$=0.02, $T_s\sim$140 K for $x$=0.04, and $T_s\sim$ 140 K for $x$=0.05, indicating the first-order character of these transitions.
Furthermore, the intensity of the superstructures in superconducting $x$=0.05 is suppressed below $T<$ 50 K again.
Similar suppression of the CDW intensities near the superconducting dome was reported for CDW in high-temperature superconducting cuprates \cite{Ghiringhelli12}, indicating that these systems may harbor similar exotic phases.

\subsection{REXS at Ir $L_3$ edge}

The spatial modulation of the electronic states in the Ir sites was investigated using the resonance at the Ir $L_3$ absorption at $h\nu\sim$ 12.2 keV \cite{Kamitani13,Clancy12}.
Figure 4 shows the REXS and x-ray absorption spectra for (a) $x$=0.0 and (b) $x$=0.04.
However, no noticeable {\bf Q}-dependence is observed for the REXS spectra at the Ir $L_3$ edge.
Although the REXS signals at the Te edges are resonantly enhanced in the XAS-peak region, as shown in Ref. \cite{Takubo14} and later sections,
only the dip structures were observed on REXS at the Ir $L_3$ edge. 
While the REXS spectra for the Te edges can be modeled by using (i) valence modulation or (ii) energy shift models in the previous study \cite{Takubo14,Achkar13},  
such dip structures on REXS without a {\bf Q}-dependence can only be reproduced by the calculation with (iii) a lattice displacement model such as given in the bottom of Fig. 4(a).
In this calculation, the form factors $f(\omega)$ for different Ir and Te sites are evaluated from XAS.
The wavevector (\textbf{Q}) and photon-energy ($\omega$) dependent structure factor $S (\mathbf{Q},\omega)$ are subsequently constructed based on the spatial modulation of $f(\omega)$ at different atomic positions $\mathbf{r}_j$:
\begin{equation} 
S (\mathbf{Q},\omega) = \sum_{j}{f_j(\omega) e^{-i \mathbf{Q} \cdot \mathbf{r}_j}}.
\end{equation}
In the lattice displacement model, the major contribution to $S (\mathbf{Q},\omega)$ originates from $\mathbf{r}_j\!=\!\mathbf{r}_j^0 + \delta \mathbf{r}_j$,
where small displacements are used for the Ir and Te lattice sites in the modulated structure.
Here, $f_j(\omega)$ are site-independent; namely, no electronic modulation is assumed in the Ir sites.
If $f_j(\omega)$ are assumed to have a modulation as large as $\sim$0.3 eV, the calculated line shape exhibits a large {\bf Q}-dependence
and conflicts with the present experimental observation (with further details of the calculation given in the Appendix).
Therefore, these REXS results indicate that the electronic states in the Ir sites scarcely have spatial modulation,
apparently contradicting the previous x-ray photoemission results \cite{Ko15,Qian13,Ootsuki12}.
Since REXS at the Ir $L_3$ edge is a highly bulk-sensitive technique compared to photoemission,
the discrepancy may arise from the electronic structural difference between the bulk and surface regions suggested in the recent STM experiment \cite{Kim16}.
The charge ordering in the Ir sites may only exist in their surface region,
but the charge ordering in the bulk of Ir$_{1-x}$Pt$_x$Te$_2$ seems to occur in the Te orbitals rather than the Ir orbitals.

\begin{figure}[ht]
	\includegraphics[width=1\linewidth]{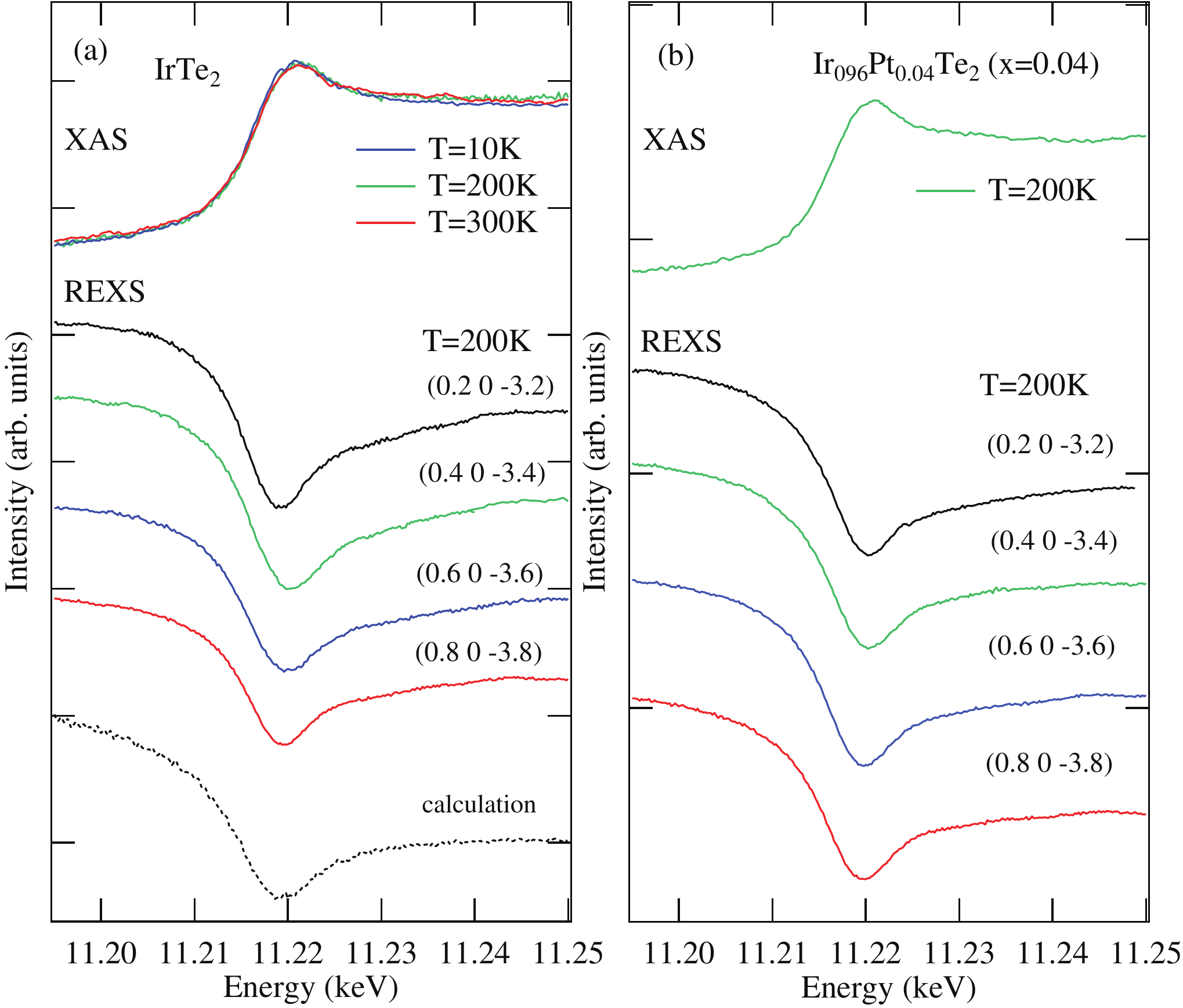}
	\caption{
		(color online.) REXS and XAS spectra at the Ir $L_3$ edges for (a) IrTe$_2$ ($x$=0.0) and (b) Ir$_{0.96}$Pt$_{0.04}$Te$_2$ ($x$=0.04).
		The calculated spectrum using the lattice displacement model (see main text).
	}
\end{figure}

\subsection{RIXS at Ir $L_3$ edge}

To further examine the electronic structural evolution in the Ir sites at the specific {\bf Q} positions,
RIXS spectra have been acquired with incident energies of $h\nu_i$ = 11.214 keV near the top of the Ir $L_3$-edge.
The data are shown in Fig. 5(a) and 5(b) for IrTe$_2$ and in Fig. 5(c) for Ir$_{0.95}$Pt$_{0.05}$Te$_2$ ($x$=0.05).
Distinct elastic diffraction ($E_{loss}$=0) is observed at the superstructural positions of {\bf Q}=(1.4 0 6.6) and (1.6 0 6.4) at $T$=200 K (below $T_s$) for IrTe$_2$.
Asymmetric lineshape of the elastic peak comes from residual strain in the diced analyzers crystal.
Some diffusive diffraction along ($h$ 0 $-h$) is also observable at {\bf Q}=(1.7 0 6.3) at $T$= 200 K.
Strong fluorescence is observed around 2-4 eV energy loss at all positions, which is associated with the hybridization effects between the transition-metal $d$ and chalcogen $p$ states \cite{Monney13,Takubo17}.
In contrast, the $d$-$d$ excitations across the Ir $t_{2g}$ bands observed between 0.5 and 1.5 eV are very weak compared to those obtained for Ir oxides \cite{Ishii13}.
These observations indicate that the holes near the Fermi-level and its spatial modulation reside in the Te orbitals rather than in the Ir orbitals and are consistent with the REXS results at the $L_3$ edge described before.
Although the spectral change across $T_s$ is very small, the tendency is similar to that observed in RIXS for CuIr$_2$S$_4$ across $T_{MIT}$ [see averaged (sum) spectrum in Fig. 5(c) and in Fig. 4 of Ref. \onlinecite{Gretarsson11}].
The spectral weight near the Fermi-level ($E_{loss}\sim$0.5 eV) are transferred into the higher energy region of $\sim$1.5 eV below $T_s$, qualitatively in agreement with the reconstruction of band structure near the Fermi level up to 2 eV in the optical conductivity measurement \cite{Fang13}.
In addition, the spectral difference between Ir$_{0.95}$Pt$_{0.05}$Te$_2$ ($x$=0.05) and the low temperature phase of IrTe$_2$ taken at $T$=200 K is similar to the temperature dependence of IrTe$_2$ across $T_s$ [Fig.5(c)],
indicating that these electronic evolutions definitely reflects the structural transition of these systems.
The spectral shapes of the $d$-$d$ excitation scarcely depend on the {\bf Q} positions, which are also similar to the case of CuIr$_2$S$_4$ \cite{Gretarsson11}.

\begin{figure}[t]
	\includegraphics[width=1\linewidth]{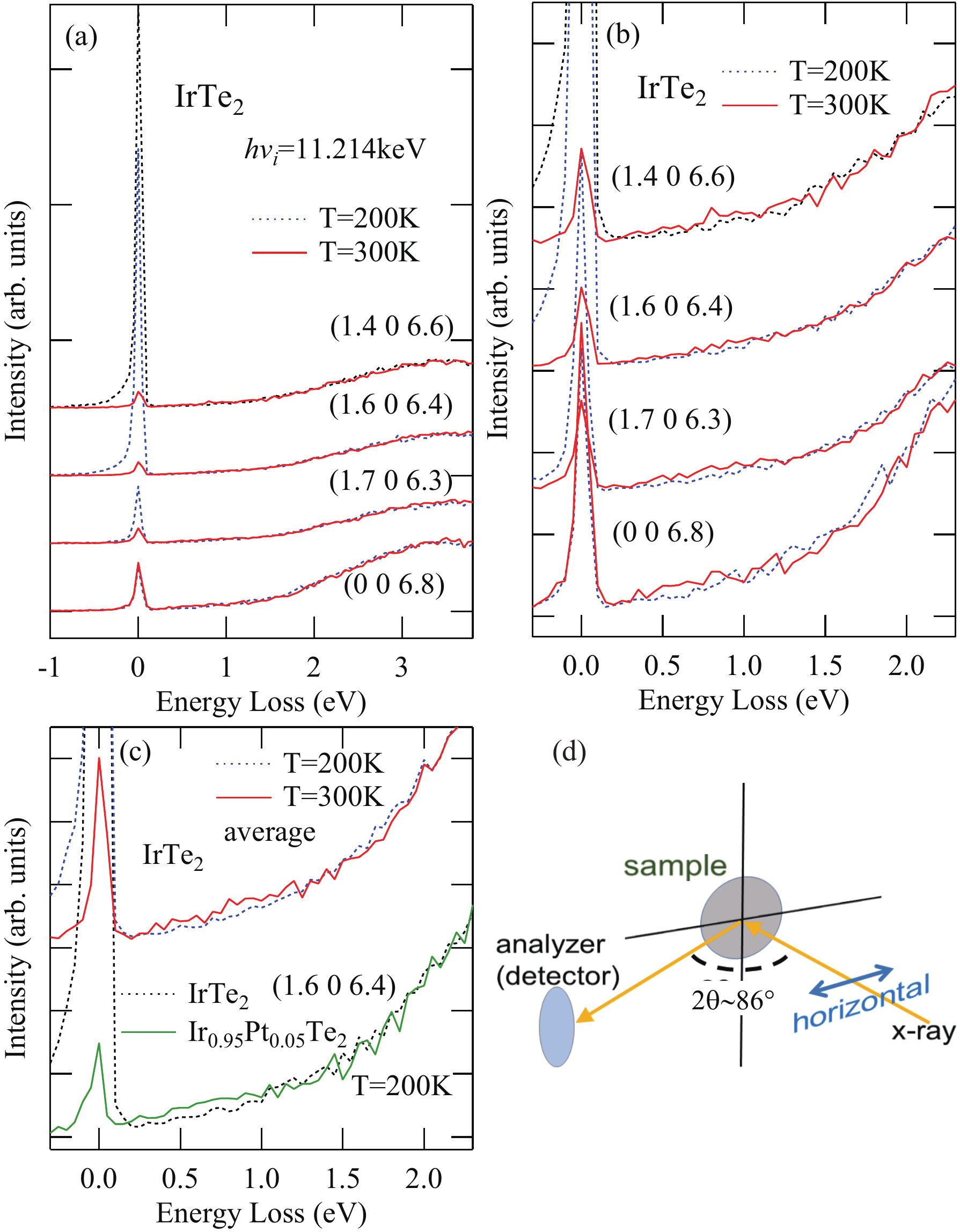}
	\caption{
	(color online.) RIXS of IrTe$_2$ and Ir$_{0.95}$Pt$_{0.05}$Te$_2$ with incident x-ray energies of $hv_i=11.214$ keV at the Ir $L_3$ edges.
	(a) RIXS for IrTe$_2$ at the selected {\bf Q} positions.
	(b) Enlarged view of RIXS in the low-energy-loss region.
	(c) Averaged spectra for IrTe$_2$ at $T$=200 and 300 K, which are taken as an average of the spectra
	shown in (b) (upper).
	RIXS on {\bf Q}=(1.6 0 6.4) for IrTe$_2$ and Ir$_{0.95}$Pt$_{0.05}$Te$_2$ at $T$=200K (lower).
	(d) Experimental geometry of RIXS.
	}
\end{figure}

\subsection{REXS at Te $M_5$ and $L_1$ edges}

Finally, the spatial modulation in the Te sites is investigated using REXS at the Te edges.
Figure 6 shows ($H$ 0 $L$) momentum scans and their temperature dependences through the resonant peak for Ir$_{1-x}$Pt$_x$Te$_2$ ($x$=0.0, 0.04 and 0.05) at a photon energy of 571.3 eV, corresponding to the Te-$M_5$ pre-peak position.
REXS signals on the superstructures around {\bf Q}=(0.2 0 $-$0.2) are clearly observed on all samples of $x$=0.0, 0.04, and 0.05 at the low temperature, consistent with the hard x-ray experiments described before.  
The CDW incomensurations, namely peak-shift to the lower-$H$ side and broadening of the widths are also found on $x$=0.04 and 0.05 near the onset of the superconducting dome.
The cooling rates were $\sim$ 4 K/min. and the cascade transition to ${\mathbf Q}_{1/8}$ was not found in these soft x-ray experiments.
This difference could arise from the difference of the charge order between the surface and bulk region due to the different penetration depths for soft and hard x-ray experiments, while {\bf Q}$_{1/8}$-type order was reported on a surface sensitive STM study by Hsu in Ref. [27].
Similar to the hard x-ray experiment, all of the superstructural signals show sharp onsets at $T\sim$ 280 K for $x$=0.0, 130 K for $x$=0.04, and 120 K for $x$=0.05, respectively,
indicating again the first-order character of these phase transitions.

\begin{figure}[htb!]
	\includegraphics[width=1\linewidth]{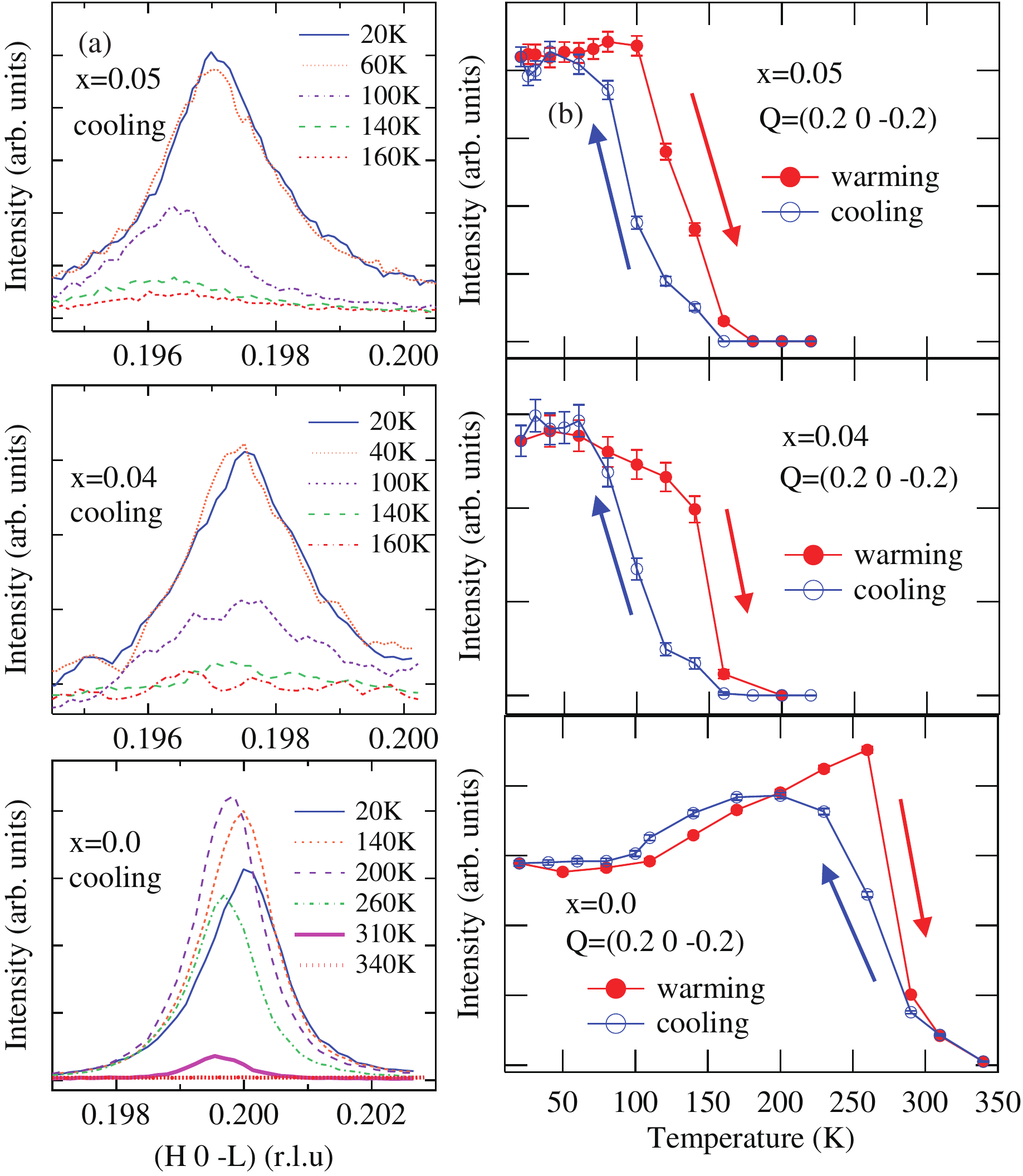}
	\caption{(color online). Temperature dependence of REXS at the Te $M_5$ edges with 571.5 eV photons.
		(a) REXS ($H$,0,-$L$) scan through the $\mathbf{Q}$=(1/5 0 $-$1/5) superstructure peak measured on Ir$_{1-x}$Pt$_x$Te$_2$ [$x$=0.0(bottom), $x$=0.04(middle), $x$=0.05(top)].
		(b) Corresponding temperature dependence of the REXS intensity,
		which were evaluated as the sum of the counts over the whole peak of (0.2 0 $-$0.2).
	}
\end{figure}

Figure 7 shows the Te-$M_5$ pre-edge spectra which reflect the covalency between Te 5$p$ and Ir 5$d$ orbitals, or unoccupied partial density of states (DOS) in the Te sites \cite{Takubo14}.
The XAS spectra at $T$=300 K shift to higher energy as the doping $x$ increases, indicating chemical potential shifts or electron dopings into the Te 5$p$ orbitals induced by the Pt substitution [Fig.7 (a)].
The spectral change of XAS across the transition for IrTe$_2$ is consistent with the result of the band structure calculations \cite{Pascut13,Toriyama13,Qian13}.
While the XAS spectra barely show any temperature dependence for $x$=0.05, the energy-dependent lineshape of REXS for $x$=0.05 at $T$=20 K is very similar to that for $x$=0.0,
as can be seen in Fig. 7(b) and 7(c).
These features of REXS for $x$=0.0, namely dip-hump structures at Te-$M$ is an evidence of the modulation of the unoccupied DOS for the five structurally inequivalent Te sites \cite{Takubo14}.
The charge ordering in the Te sites of $x$=0.05 will exist in the partial small domain and 
the spatial modulation in it will be qualitatively similar to that for $x$=0.0.

\begin{figure}[thb!]
	\includegraphics[width=0.95\linewidth]{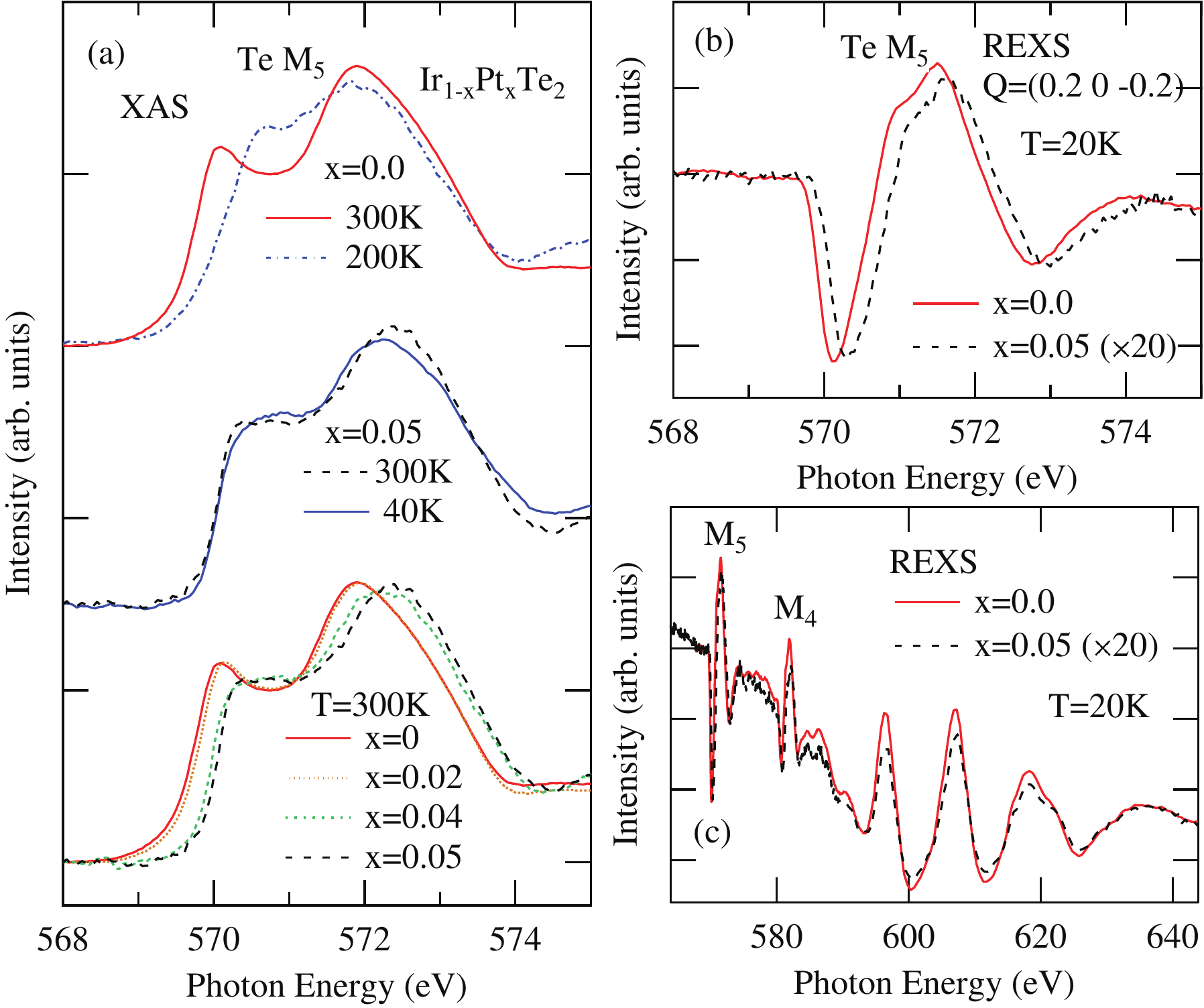}
	\caption{(color online). Comparison between XAS and REXS spectra of the Te-$M$ edge x-ray absorption for Ir$_{1-x}$Pt$_{x}$Te$_2$. (a) XAS spectra in the Te $M_5$ pre-edge region. Spectra for $x$=0.0 (top), $x$=0.05 (middle), and various compositions at $T$=300 K (bottom).
		(b) REXS spectra in the Te $M_5$ pre-edge region for $x$=0.0 and $x$=0.05 at $T$=20K.
		(c) REXS spectra in the entire energy range of the Te-$M$ edge at 20 K.}
\end{figure}

One may consider that REXS at the Te $M_{4,5}$ edges is a rather surface sensitive technique comparable to the photoemission spectroscopy,
since the Te $M_{4,5}$ edges are in the soft x-ray region. 
Then the fact that the charge modulation on REXS are observed at Te $M_{4,5}$ but not at Ir $L_3$ edge will not reflect the difference between the Te and Ir sites.      
However, the dip-hump structure and {\bf Q} dependent lineshape are also observed on the bulk-sensitive REXS at the Te $L_1$ edge as shown in Fig. 8.
The XAS and REXS spectra at the Te $L_1$ edge for IrTe$_2$ are plotted in Fig. 8(a) and a corresponding momentum scan is plotted in Fig. 8(b). 
The signals of REXS on {\bf Q}=(0.2 0 $-$2.2) and (0.4 0 $-$2.4) are resonantly enhanced at the Te $L_1$ edge and the dip features are observed before the peak structures, which is just similar to REXS at the Te $M_5$ edge.
In addition, REXS at the Te $L_1$ edge exhibits a certain {\bf Q} dependence in contrast to the case for the Ir $L_3$ edge.
The resonant enhancement at the $L_1$ peak of {\bf Q}=(0.4 0 $-$2.4) is indistinctive compared to that of {\bf Q}=(0.2 0 $-$2.2), while the dip structures are noticeable at the both positions. 
These line shapes for IrTe$_2$ can be modeled by the valence-modulation model with just same parameters for REXS at the Te $M$ edges given in Ref. \cite{Takubo14} [Fig. 8(c)].
In the valence-modulation model, the major contribution to $S ({\mathbf Q},\omega)$ arises from 
$f_j(\omega)\!=\! f(\omega,p+\delta p_j)$, where $\delta p_j$ is the variation in the local valence of the Te ions.
The parameters are $\delta p_2$ =-0.6, $\delta p_3$ = -0.6, $\delta p_5$ = -0.15, $\delta p_1$ =0.9, and $\delta p_4$ =-0.15 
and proportional to the local DOS at the Te sites which are illustrated in Fig.1 (a).
Therefore, we safely conclude that the spatial charge modulation of IrTe$_2$ exists in the Te sites even in the bulk region, following the striped formation with ${\mathbf Q}_{1/5}$=(1/5 0 $-$1.5).

\begin{figure}[htb!]
	\includegraphics[width=1\linewidth]{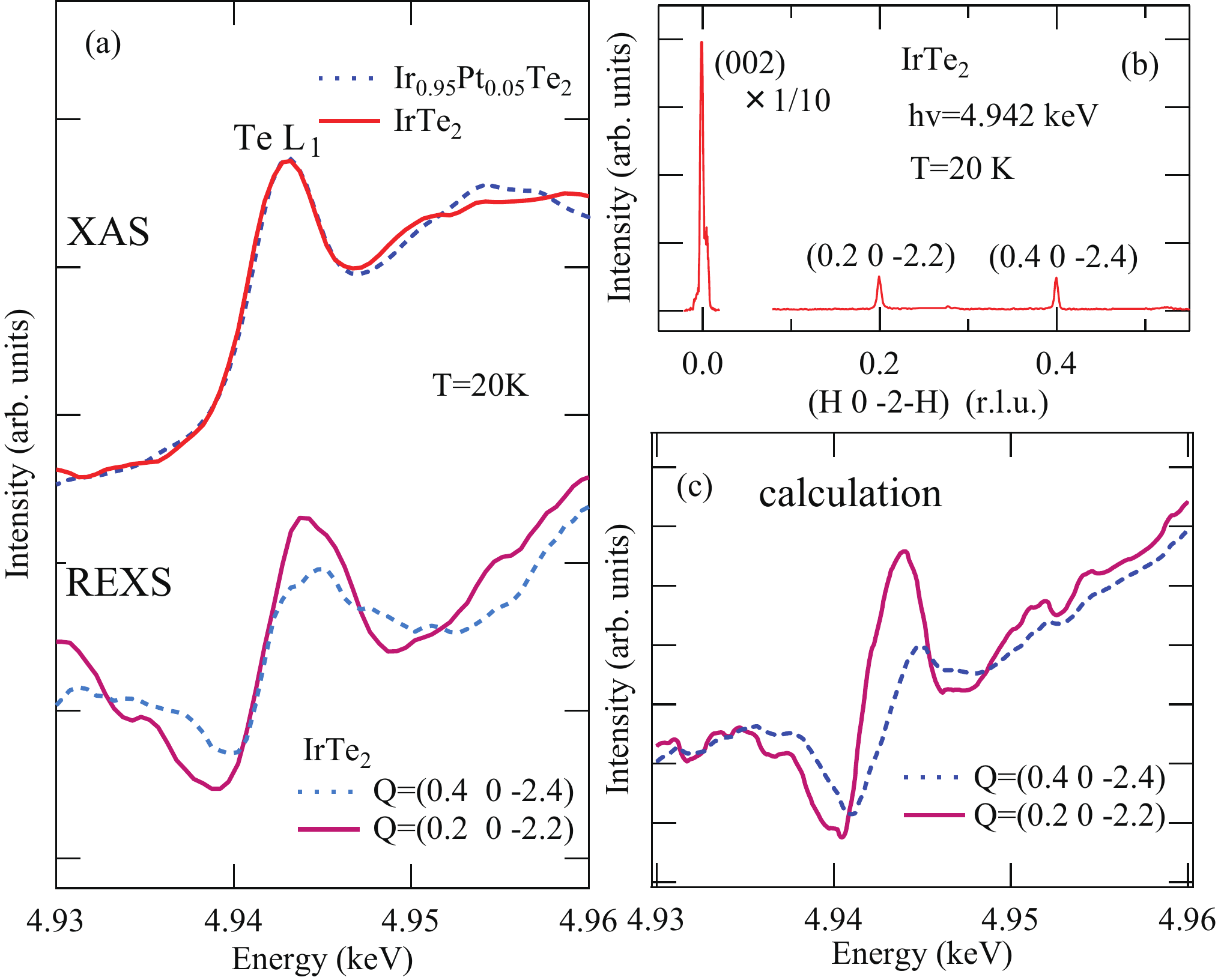}
	\caption{(color online).
		(a) Comparison between XAS and REXS spectra of the Te-$L_1$ absorption edge for Ir$_{1-x}$Pt$_{x}$Te$_2$.
		(b) Corresponding REXS ($H$ 0 $L$) scan through the superstructure peaks at $T$=20 K.	
		(c) Calculated REXS intensity for the combination of a valence-modulation model (resonant term) with nonresonant
		lattice displacements for {\bf Q}=(0.2 0 -2.2) and {\bf Q}=(0.4 0 -2.4) (see discussion in the main text.) 
	}
\end{figure}

\section{Summary}
We have examined the charge modulation of the Ir and Te sites in Ir$_{1-x}$Pt$_x$Te$_2$ by means of the resonant x-ray scattering technique.
The {\bf Q}=(1/5 0 $-$1/5)-, (1/8 0 $-$1/8)-, and (1/6 0 $-$1/6)- type superstructures are observed for IrTe$_2$ ($x$=0.0) at low temperature.
The superstructures around ${\mathbf Q}_{1/5}$=(1/5 0 $-$1/5) coexist with the superconducting phase for Ir$_{1-x}$Pt$_x$Te$_2$ ($x$=0.05), suggesting CDW persist to higher Pt substitution than previously thought.
The incommensuration of CDW is observed for $x$=0.04 and 0.05 samples which coincide with the onset of
the superconductivity.
The REXS and RIXS spectra for the Ir $L_3$ edge scarcely depend on the wavevectors,
while REXS spectra at the Te edges indicates the spatial charge modulation on the Te sites.
The charge modulation in the bulk regions seems to reside in the Te orbitals rather than the Ir orbitals.

\section*{Acknowledgements}

The authors thank Dr. D. Ootsuki for valuable discussions.
REXS measurement at the Te $M_5$ edges in this paper was performed at the Canadian Light Source, which is supported by the Canada Foundation for Innovation, Natural Sciences and Engineering Research Council of Canada, the University of Saskatchewan, the Government of Saskatchewan, Western Economic Diversification Canada, the National Research Council Canada, and the Canadian Institutes of Health Research.
REXS measurements at the Ir $L_3$ edge were performed under the approval of the Photon Factory Program Advisory Committee (Proposals No. 2015G556, 2015S2-007) at the Institute of Material Structure Science, KEK.
RIXS measurement at the Ir $L_3$ edge and REXS measurements at the Te $L_1$ edge were performed under the approval of SPring-8 (Proposals No. 2014B3787 and 2016A3564).
This works was supported by the Japan Society for the Promotion of Science (JSPS) of Grant-in-Aid for Young Scientists (B) (No. 16K20997).

\appendix

\section{Detail of the calculation of REXS intensity}

The calculation of the REXS intensity is structured similarly to the method in the previous REXS study for IrTe$_2$ at the Te $M$ edges \cite{Takubo14}.
The details were presented in the supplementary of Ref. \onlinecite{Takubo14}, thus the essential parts are pointed here. 
The scattering intensity can be expressed as:
\begin{equation}
I_\mathrm{REXS} = \frac{C|S(\bm{Q},\omega)|^2}{\mu(\omega)}.
\end{equation}
The calculation is performed for three different methods \cite{Achkar13}, namely: (i) \textit{valence modulation model}, corresponding to a periodic variation in the local valence of Ir or Te ions; (ii) \textit{energy shift model}, assuming a spatial modulation in the energy of the Ir 5$d$ or Te 5$p$ states; and (iii) \textit{lattice displacement model}, where small displacement are used for the Te and Ir lattice sites in the supermodulated structure.
These models are subsequently implemented in the calculation of the structure factor $S(\bm{Q},\omega)$ which is written generally as Eq. (1) in the main text.
In Eq. (1), $r_j^0$ is the position vector in the undistorted structure at site $j$ and $\delta r_j$ is the displacement from the lattice position due to the structural modulation.
The atomic form factor can also depend on additional parameters related to the electronic structure of the atom at $j$, such as the local charge density or energy levels; these factors are explicitly included in the respective models.
More specifically, all the energy dependent terms are included in the atomic form factor $f_j(\omega)$, while the atomic positions or displacements are of course energy-independent.

The form factor Im$\{f_j(\hbar\omega)\}$ can be determined from the XAS spectra which are offset and scaled to calculated values of the absorption coefficient $\mu (\omega)$ (from NIST \cite{NIST95}) in order to express $\mu (\omega)$ in units of $\mu m^{-1}$.
Via the optical theorem, Im$\{f_j(\omega)\}$ is linearly proportional to the absorption coefficient  $\mu (\omega)$, and Re$\{f_j(\omega)\}$ can be determined from Im$\{f_j(\omega)\}$ using Kramers-Kronig transformations.
Accordingly, to express $f(\omega)$ in electrons/atom, experimental XAS have been scaled and extrapolated to high and low energy using tabulated calculations of Im$\{f_j(\omega)\}$ above and below the absorption edge.

\subsection{Valence modulation model}
The valence (or local DOS) modulation model takes into account spatial modulations in the local DOS at the Ir and Te sites as; 
\begin{equation} 
S = \sum_{j=1}^{5}{\left[\exp\left(\frac{-2j\pi i}{5}\right) f(\omega, p + \delta p_j)\right]}.
\end{equation}

We determine $f(\omega, p_j)$ as a function of the local DOS modulation $p_j$ by performing a linear extrapolation from the $f(\omega)$ measured with XAS at 300 K (HT phase) and 200 K (LT phase).
The scattering intensity is given by:
\begin{equation}
I= \frac{C}{\mu(\omega)} \left|\sum_{j=1}^{5}{\left[\exp\left(\frac{-2j\pi i}{5}\right)f(\omega, p_j)\right]} \right|^2.
\end{equation}

The corresponding calculation for the Ir $L_3$ edge, shown in Fig. 9(a), uses parameters $\delta p_1$ =-0.6, $\delta p_2$ = -0.6, $\delta p_3$ = -0.15, $\delta p_4$ =0.9, and $\delta p_5$ =-0.15 which were used in the previous study.
However, the results have a certain $Q$-dependence and cannot reproduce the experimental feature for the Ir $L_3$ edge. 
On the other hand, these models with nonresonant lattice displacement terms describe well the REXS results at the Te- $M$ in the previous study \cite{Takubo14} and $L_1$ edges as discussed in Sec. III.E. 

\begin{figure}[t!]
	\includegraphics[width=\linewidth]{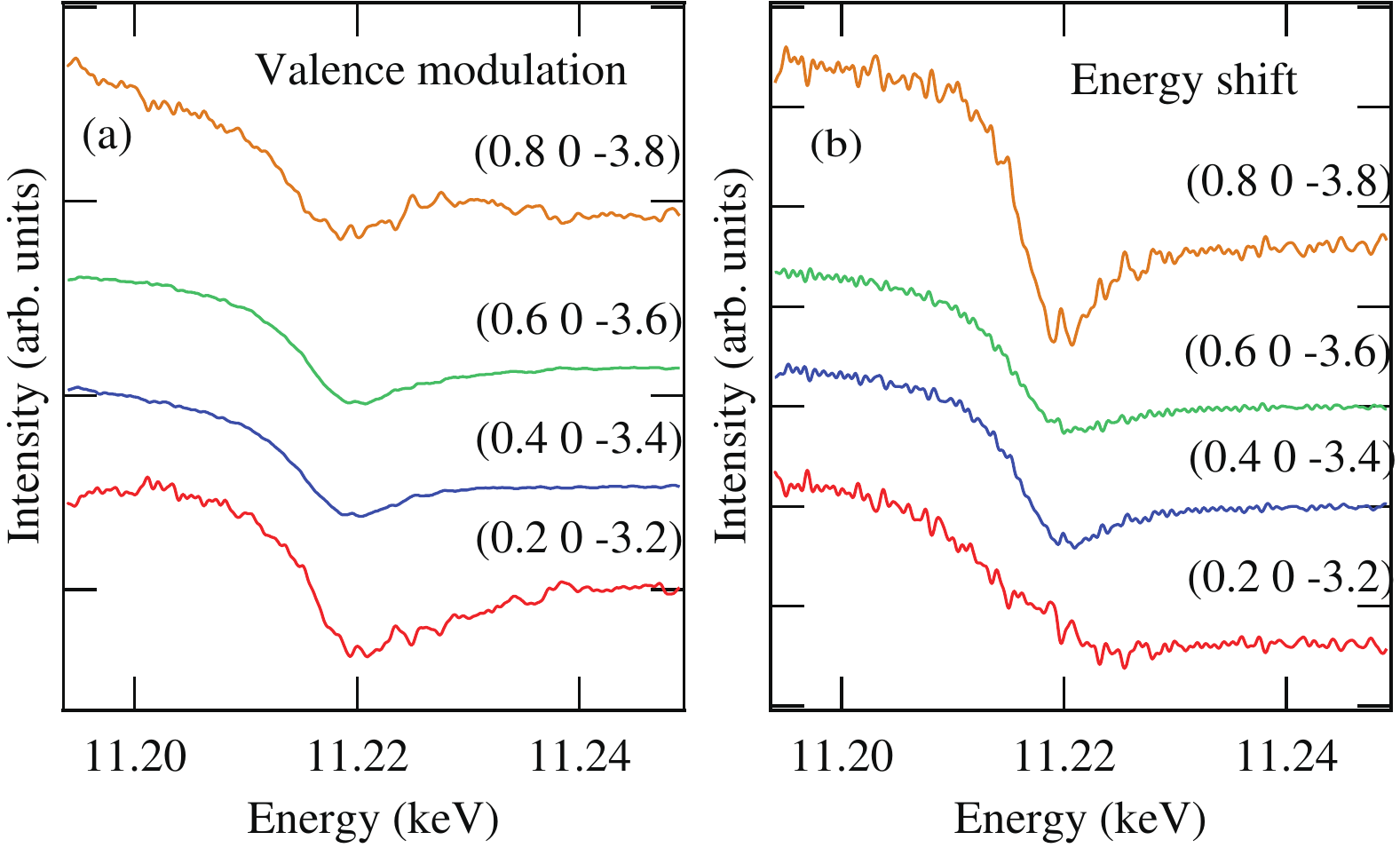} %
	\caption{Calculated REXS intensity using the (a) valence modulation and (b) energy shift models at the Ir $L_3$ edge.}
\end{figure}

\subsection{Energy shift model}
Secondly, the enegy shift model is considered.
The difference with the valence modulation model is that here we use the spatial variation of the energy shift in place of the local DOS; in this case, the structure factor for the Ir-striped model is given by:
\begin{eqnarray}
S &=& \sum_{j=1}^{5}{\left[\exp\left(\frac{-2j\pi i}{5}\right)f(\hbar\omega+ \Delta E_j)\right]}. \\
\end{eqnarray}
Therefore the scattering intensity can be written as:
\begin{equation}
I= \frac{C}{\mu(\omega)} \left|\sum_{j=1}^{5}{\left[\exp\left(\frac{-2j\pi i}{5}\right)f(\hbar\omega+ \Delta E_j)\right]} \right|^2.
\end{equation}

The calculated results for Ir $L_3$ using this model are shown in Fig. 9 (b)
with $\Delta E_1$ = -0.3 eV, $\Delta E_2$ = -0.05 eV, $\Delta E_3$ = 0.2 eV, $\Delta E_4$ = 0.2 eV, $\Delta E_5$ = -0.05 eV, which parameters were used in the previous study for the Te $M_5$ edge.
However, the {\bf Q}-dependence are appeared on this calculation similar to the valence modulation model, and therefore REXS at the Ir $L_3$ edge cannot be described within this model.

\subsection{Lattice displacement model}
For the lattice displacement model,
$f_j$ is the same at each site, but lattice positions are displaced, i.e. $\bm{r_j}=\bm{r_j}^0 + \delta \bm{r_j}$.
Considering a chain of 5 Te and/or Ir sites separated by ($a_H$, 0, -$c_H$) $\sim$ (0, 0, $c_L/5$), the structure factor is given by:
\begin{equation}
S = \sum_{j=1}^{5}{\left[\exp\left(\frac{-2j\pi i}{5}+\frac{\delta_j}{c_L}\right)\right]}f(\omega).
\end{equation}
In the limit of small displacements, we can expand the exponential terms to first order and write the REXS intesity as:
\begin{eqnarray}
I &\cong& \frac{4\pi^2C}{\mu(\omega)c_L^2}\left| {\sum_{j=1}^{5}}\left[\exp\left(\frac{-2j\pi i}{5}\right)\delta_j\right]f(\omega)\right|^2 \\ 
&\propto&\frac{|f(\omega)|^2}{\mu(\omega)}.
\end{eqnarray}
This result holds even if one includes higher order terms in the series expansion.
Moreover, the magnitude of the displacements has no impact on the energy nor {\bf Q}-dependence of the calculated scattering intensity, since no modulation was assumed in the electronic state of each sites or $f(\omega)$.
The calculated result for the Ir $L_3$ edge using lattice displacement model are plotted in the bottom of Fig. 4 (a)
and well reproduced the experimental result.
On the other hand, the REXS at Te $L_1$ edge are unable to be reproduced by this model, since the experimental
results exhibits a certain {\bf Q}-dependence as discussed in the main text.


\end{document}